\documentclass[prd,aps,eqsecnum,floatfix,nofootinbib,preprint,tightenlines]{revtex4}
\usepackage{latexsym}
\usepackage{graphicx}
\usepackage{amsmath}
\usepackage{hyperref}

\def\NON{\nonumber\\}
\def\bibi{\bibitem}

\def\d{\delta}
\def\e{\epsilon}                
\def\g{\gamma}

\def\l{\lambda}

\def\p{\pi}                     

\def\D{\Delta}

\def\X{\Xi}

\def\ca{{\cal A}}

\def\cd{{\cal D}}

\def\co{{\cal O}}

\def\cbo{{\,\raise-.15ex\Sc [\,}}                       

\def\svev#1{\left\langle #1\right\rangle}       

\def\ddt#1{{\buildrel {\hbox{\LARGE .\kern-2pt.}} \over {#1}}}

\def\ie{\mbox{\it i.e.} }
\def\eg{\mbox{e.g.} }

\def\tr{{\rm tr}\,}
\def\Tr{{\rm Tr}\,}

\def\Det{{\rm Det}}

\def\ie{{\it i.e.}}

\def\eg{{\it e.g.}}

\def\et{{\it et al.}}

\def\Ibar{{\overline{I}}}
\def\qbar{{\overline{q}}}
\def\psibar{{\overline{\psi}}}
\def\phibar{{\overline{\phi}}}
\def\etabar{{\overline{\eta}}}

\def\tJ{{\tilde J}}
\def\Det{{\rm Det}}
\def\Tr{{\rm Tr}}
\def\tr{{\rm tr}}
\def\Exp{{\rm exp}}
\def\I{{\bf 1}}

\def\seneqs#1#2{(\ref{#1}) and (\ref{#2})}

\begin{document}
\hyphenation{fer-mio-nic per-tur-ba-tive pa-ra-me-tri-za-tion
pa-ra-me-tri-zed a-nom-al-ous}

\renewcommand{\thefootnote}{$*$}

\begin{center}
\vspace*{10mm}
{\large\bf 't~Hooft vertices, partial quenching, \\[4mm]
  and rooted staggered QCD}
\\[12mm]
Claude Bernard,$^a$ Maarten Golterman,$^b$ Yigal Shamir$^c$\ and\ Stephen R.\ Sharpe$^d$
\\[8mm]
{\small\it
$^a$Department of Physics
\\Washington University,
Saint Louis, MO 63130, USA}
\\[5mm]
{\small\it
$^b$Department of Physics and Astronomy
\\San Francisco State University,
San Francisco, CA 94132, USA}
\\[5mm]
{\small\it $^c$Raymond and Beverly Sackler School of Physics and Astronomy\\
Tel-Aviv University, Ramat~Aviv,~69978~Israel}
\\[5mm]
{\small\it $^d$Physics Department\\
University of Washington, Seattle, WA 98195, USA}
\\[10mm]
{ABSTRACT}
\\[2mm]
\end{center}

\begin{quotation}
We discuss the properties of 't~Hooft vertices
in partially quenched and rooted versions of QCD in the continuum.
These theories have a physical subspace, equivalent to ordinary
QCD,  that is contained within a larger space that includes many unphysical correlation functions.
We find that the 't~Hooft vertices in the physical subspace have the
expected form, despite the presence of unphysical 't~Hooft vertices appearing
in correlation functions that have an excess of valence quarks (or ghost quarks).
We also show that, due to the singular behavior of unphysical
correlation functions as the massless limit is approached,
order parameters for non-anomalous symmetries
can be non-vanishing in finite volume if these symmetries act
outside of the physical subspace.
Using these results,
we demonstrate that arguments recently given by Creutz---claiming to
disprove the validity of rooted staggered QCD---are incorrect.
In particular, the unphysical 't~Hooft vertices do not present an
obstacle to the recovery of taste symmetry in the continuum limit.
\end{quotation}

\renewcommand{\thefootnote}{\arabic{footnote}} \setcounter{footnote}{0}

\newpage
\section{\label{intro} Introduction}
Staggered fermions are a relatively inexpensive method for putting
quarks on the lattice, and have led to a
growing number of precise QCD predictions
that can be compared with experiment.
There is, however, a theoretical issue that needs to be understood in detail
in order to be confident that there are no hidden systematic errors, \ie,
errors that, because their nature is not understood, are not being taken into
account.

If one uses a staggered fermion field for each (light) physical
flavor, then the continuum limit of QCD with staggered fermions has
too many quark degrees of freedom.
Proceeding naively,
each staggered fermion yields four degenerate Dirac
fermions in the continuum limit.
As has become customary, we will refer to this extra multiplicity as
``taste.'' In the continuum limit the quark fields carry a taste index,
and correspondingly, have an $SU(4)$ taste symmetry.
On the lattice the $SU(4)$ taste symmetry is explicitly broken,
but the effects of this breaking
are controlled by the lattice spacing $a$. In particular,
taste breaking leads to order $a^2$ effects in physical quantities,
provided that none of the quark masses vanishes.

In order to reduce the four-fold multiplicity, for each physical
flavor one takes the fourth
root of the determinant of the staggered Dirac operator
inside the lattice path integral that defines the theory.
This ``fourth root trick'' has been used
extensively in practice.
At the level of
lattice perturbation theory, it is
straightforward to see that this indeed removes the extra degeneracy
of each closed fermion loop, thus restoring the correct number
of sea quarks \cite{bgpq,sharpelat06}.
The key question is whether the use of ``rooted'' staggered QCD
is legitimate also at the non-perturbative level.

This question has received much attention recently
(see Refs.~\cite{dh1} to \cite{Aubin}),
and there is mounting evidence,
both analytical and numerical, that the continuum
limit of rooted staggered QCD is in the correct universality class.\footnote{%
The concerns raised in Ref.~\cite{creutz1} have been
answered in Ref.~\cite{bgss}. The additional concerns brought up in
Refs.~\cite{creutz2,creutz} are answered here.}
For details, we refer the reader to the original literature
as well as to the review articles Refs.~\cite{sharpelat06,bgslat06,kronfeldlat07}.
In a nut-shell,  the combined effect of lattice taste-breaking and
the fourth root trick makes rooted staggered QCD a non-local
lattice theory.
In a careful treatment of the continuum limit, in which all assumptions
have been spelled out, these effects have been argued
to vanish, with the conclusion that
the set of physical correlation functions of the rooted theory
is reproduced by a non-perturbatively well-defined, local theory,
provided that the continuum limit is taken before the chiral limit.
The chiral effective theory that reproduces the light pseudoscalar
sector of rooted staggered QCD, including its non-local discretization
effects, has been shown to be staggered chiral perturbation theory
with the replica trick.
These arguments imply that $SU(4)$ taste symmetry is restored in the continuum limit.

The corresponding continuum theory therefore has
four copies (``tastes'') of each flavor, but with the $1/4$ power of
the determinant appearing in the functional integral.  We call such a rooted
theory with exact taste symmetry
a ``rooted continuum theory'' (RCT).
It is easy to see \cite{bgss} that
the physical subspace of the RCT reproduces the correlation functions of QCD.
The relevant discussion from Ref.~\cite{bgss} is summarized below.  We emphasize
here that the equivalence between QCD and the physical subspace of
the RCT is
rigorously established (as long as the quark masses are positive).
What is less certain (though very likely, we believe,
based on Refs.~\cite{sharpelat06,cbrschpt,shamirrg,bgseft})
is that the rooted staggered theory on the lattice becomes this
RCT as $a\to0$.

Despite having a physical subspace equivalent to QCD,
the RCT is not identical to QCD.
The presence of four
tastes for each physical flavor allows one to construct additional, unphysical, correlation functions
beyond the physical set of correlations that occur in QCD.
When taste symmetry is exact, the fourth root removes three of the
four tastes from the quark sea for each physical flavor,
and there is thus an ``oversupply'' of valence quarks.
This can be restated:  a rooted theory with exact taste symmetry is
precisely equivalent to  a partially quenched theory \cite{bgpq,cbrschpt,sharpelat06},
with four normal-statistics quarks (the four tastes)
and three opposite-statistics quarks, or ghost quarks,
for each flavor.

The question then is what type of projection or
averaging should be used in order to construct correlation functions
that correspond to the physical subsector of the partially quenched theory.
We emphasize that this question is of
practical importance.  In order to extract the physical
observables of QCD we need to know which correlation
functions in the rooted lattice theory become,
in the continuum limit, equal to correlation
functions in the physical subsector.
The resolution of this ``valence'' problem of the RCT
is much simpler than the question of whether the rooted staggered
theory has the expected continuum limit in the first place.  It is in fact
possible to solve the valence problem completely,
and a general analysis has already been given in Refs.~\cite{bgss,sharpelat06}.

Here, we will revisit the valence problem,
focusing on the role of 't~Hooft vertices in rooted and/or partially quenched versions
of  QCD.
More precisely, we consider QCD with one flavor, where the issues are
particularly acute. The conclusions generalize straightforwardly to
multiple flavors.
Although our discussion is an application of the general analysis
of Refs.~\cite{bgss,sharpelat06}, we think it useful to describe it
explicitly as it brings out several unusual features of partially quenched theories.
Furthermore, a recent article by Creutz claims to show
that rooting fails using arguments based on 't~Hooft vertices~\cite{creutz}.
One application of our discussion here is to show that the arguments
of Ref.~\cite{creutz} are incorrect.
Indeed, Creutz's main arguments apply to the assumed continuum-limit theory with
exact taste symmetry, the RCT.  Therefore it is possible to give a complete
resolution of the apparent paradoxes that he finds.

\bigskip
The paper is organized as follows.
In Sec.~\ref{correlators}
we review in some detail the RCT
of one-flavor rooted staggered QCD. We show how the exact taste
symmetry allows one to establish rigorously (for positive quark mass)
that a rooted theory with four tastes has
a physical subspace identical
to that of standard one-flavor QCD. We also write the RCT
as an unrooted, but partially quenched, theory, which
will be particularly useful for discussing the 't~Hooft vertices.
We discuss how one projects onto the physical subsector of that theory.
We focus on the method of projection that is most useful for
the subsequent discussion, namely that based on picking out a single taste.
In Sec.~\ref{'t Hooft} we study 't~Hooft vertices in
the RCT,
showing that infrared
divergences are present in unphysical, but never in physical,
correlation functions.
In Sec.~\ref{wi} we resolve an apparent inconsistency between
the anomalous $U(1)$ chiral symmetry of the one-flavor theory
and the existence of non-anomalous chiral symmetries
that act on the enlarged set of correlation functions
of the partially quenched theory.
As an application, we show, in Sec.~\ref{MC},
how the arguments about 't~Hooft vertices in Ref.~\cite{creutz} fail. We also
discuss claims by Creutz that the RCT cannot be the
limit of the lattice theory, because (1) the tastes of the lattice theory
have ``canceling'' chiralities, while the tastes of the RCT do not, and (2) the
eigenvalue flow is an obstruction to the restoration of taste symmetry.
We show that both these claims are false, and are
based on a misunderstanding of the nature of chiral symmetry on the lattice.
We conclude in Sec.~\ref{conclusion}.

An appendix describes how the same conclusions can be reached
using the alternative method of projecting onto physical
mesonic correlation functions used in our earlier paper~\cite{bgss}.
The method does not invoke the partially quenched framework.

\bigskip
We close the introduction by emphasizing five points so as
to avoid later confusion.
The first is that all the partially quenched theories that we discuss have equal valence
and sea quark masses. For such theories one can project
onto the physical subsector using valence quarks alone (since valence and
sea quarks are interchangeable).
By contrast, most applications of partial quenching have differing
valence and sea quark masses. For such theories, all correlation functions
composed of valence quarks are unphysical, exhibiting, for example, double poles.

The second point concerns the order of the chiral and continuum limits.
All the arguments for the correctness of the universality class of rooted
staggered QCD are predicated on sending $a\to 0$ for fixed, positive,
non-vanishing quark mass.\footnote{The analytic continuation to Minkowski
  space must also be postponed until after the continuum limit.}
  The chiral limit can be taken only
after taking the continuum limit. Taking the limits in the other
order leads to incorrect answers (\eg\ the vanishing
of the condensate in one-flavor QCD). This point has been discussed
in our earlier paper~\cite{bgss} and
elsewhere~\cite{sharpelat06,shamirrg,dh1,dh2,bgslat06,cblimits}.
We assume throughout this paper that the order of limits has been taken correctly,
so that the RCT has positive quark mass, which may
only be taken to zero as a final step.

Third, a claim in Ref.~\cite{creutz}
is that 't~Hooft vertices obstruct the recovery of taste
symmetry because they lead to a non-perturbative
coupling between different tastes that survives in the continuum limit.
One might therefore worry
that we are by-passing the main issue by basing our discussion on
the RCT, which has exact vector taste symmetry from the outset.
This is not a problem because
the arguments of  Ref.~\cite{creutz} about 't~Hooft vertices in fact
respect taste symmetry, and can be considered equally well in the RCT
(or in its equivalent partially quenched version).
Put differently, we show, in the RCT, that the
non-perturbative couplings between different tastes
do not lead to any violations of vector taste symmetry or
to disagreement with QCD in the physical sector;
if the arguments of Ref.~\cite{creutz} were correct, we
would not be able to show this.
For a rooted staggered theory at non-zero lattice spacing,
there will be unphysical effects in all sectors, but since these vanish in the physical sector
of the RCT, they will be of order $a^2$ in quantities
whose continuum limit is in that sector.

Fourth, we note that the RCT can be obtained in a rigorous
way by regulating the rooted
theory in
a taste and chirally invariant fashion---for example,
by starting with four identical copies of overlap quarks---and
then taking the continuum limit.

Finally, we stress that showing the arguments of
Ref.~\cite{creutz} to be incorrect does {\em not}\/ imply that the rooted staggered
theory at non-zero lattice spacing has the
correct limit as $a\to 0$. It only shows that the assumed limiting theory
(the RCT) is not
inconsistent, and has a physical subspace equivalent to QCD.
Whether the RCT is actually obtained from the lattice theory
as $a\to0$ is a different issue, the status of which we reviewed briefly above.

\section{\label{correlators} Rooted correlation functions in the continuum limit}
We begin with a review of  relevant aspects of the valence problem
as discussed, in particular, in Appendix B of Ref.~\cite{sharpelat06}
and Sec.~3 of Ref.~\cite{bgss}.
We restrict the discussion to the one-flavor theory,
where the lattice path integral contains the positive fourth root
of the determinant of a single staggered
Dirac operator $D_{stag}+m_{stag}$.
The corresponding RCT has, by fiat, exact taste symmetry,
so its Dirac operator is given by
\begin{equation}
\label{conlim}
D_{RCT} + M =  (D+ m) \otimes\I\ ,
\end{equation}
where $D+m$ is the continuum single-quark Dirac operator,
and $\I$ is the $4\times 4$ unit matrix in taste space.
We assume a strictly positive quark mass, $m>0$. In the continuum,
the determinant of $D+m$ is then (formally) positive;  the determinant
is rigorously positive for overlap quarks.\footnote{%
Since the staggered mass
$m_{stag}$  is multiplicatively renormalized \cite{gs}, a positive
continuum mass $m$ will result from a positive $m_{stag}$.
For a negative quark mass, see Refs.~\cite{dh2,bgss}.}
We thus have
\begin{equation}
\label{determinant}
\Det^{1/4}\left(D_{RCT} + M \right)= [\Det^4(D + m)]^{1/4} = \Det(D+m) \ .
\end{equation}
Note that the positivity of the determinant on the right-hand side is
crucial here, since the positive fourth root is always taken
on the left-hand side.

The generating functional for quark correlation
functions in the RCT
is thus given by
\begin{subequations}
\label{genfun}
\begin{eqnarray}
\label{genfuna}
Z(\eta,\etabar)
\!\!&=&\!\!
\int \cd\ca\;e^{-S_g}\;\Det^{1/4}\left(D_{RCT}+M\right)\;
\Exp\left\{{\sum_{i,j=1}^4\etabar_i (D_{RCT}+M)^{-1}_{ij}\eta_j}\right\}
\hspace{5ex} \\
\!\!&=&\!\! \int \cd\ca\;e^{-S_g}\;\Det\left(D+m\right)\;
\Exp\left\{{\sum_{i=1}^4\etabar_i (D+m)^{-1}\eta_i}\right\}
\label{genfunb}
\end{eqnarray}
\end{subequations}
Here $S_g$ is the gauge action,
and we have introduced sources $\eta_i$ and $\etabar_i$ for all tastes.
For rooted staggered quarks on the lattice, $D_{RCT}+M$ in Eq.~(\ref{genfuna})
can just be replaced by the staggered Dirac operator in the taste representation.
Once the taste symmetry is restored, however, we have the option of working
with the much simpler expression~(\ref{genfunb}).

The generating functional~(\ref{genfun})
is also equal to that of an $SU(4|3)$ partially quenched theory~\cite{bgpq}:
\begin{eqnarray}
Z(\eta,\etabar)
\!\!\!&=&\!\!\!
\int \cd\ca\;e^{-S_g} \int \cd\psi\cd\psibar\cd\phi\cd\phibar\;
\nonumber\\
\label{genfun3}
&& \!\!\!\times \;
\Exp\left\{{-\sum_{i=1}^4\left(\psibar_i (D+m)\psi_i+\etabar_i\psi_i+\psibar_i\eta_i\right)
-
\sum_{j=1}^3 \phi_j^\dagger (D+m)\phi_j}\right\}\!.
\hspace{3ex}
\end{eqnarray}
Here $\phi_j$ and $\phi^\dagger_j$ are three
bosonic (ghost) quarks
whose functional integral gives the required inverse powers of $\Det(D+m)$.
The ghost-quark path integral converges because $m>0$.
Note that the valence, sea and ghost quarks all have the same mass.

The  generating functional $Z(\eta,\etabar)$
exhibits the valence problem alluded to in the introduction:
Because we can have
all four valence quarks on external lines, the set of
correlation functions defined by Eq.~(\ref{genfun}) is much larger than the
set of physical correlation functions of a one-flavor theory.

The resolution of the valence problem is obvious \cite{sharpelat06}:
Thinking of, for example, the quarks
with taste index $2,3,4$ as valence quarks, and pairing them with the
three ghost quarks, makes the quark with taste index $1$ the physical
quark.  The set of all physical correlation functions will thus be generated
by operators that depend on $\psi_1$ and $\psibar_1$ only, or, equivalently,
by  taking derivatives with respect to $\eta_1$ and $\etabar_1$ with all
other sources set to zero.
Indeed, upon setting $\eta_{2,3,4}=\etabar_{2,3,4}=0$, one immediately
sees that
the generating functional of the partially quenched theory reduces to
the generating functional of the physical one-flavor theory:
\begin{eqnarray}
\label{genfun-1flavor}
Z_{1\,{\rm flavor}}(\eta_1,\etabar_1)&=& Z((\eta_1,0,0,0),(\etabar_1,0,0,0)) \nonumber \\
&=&
\int \cd\ca\;e^{-S_g}\;\Det\left(D+m\right)\;
\Exp\left\{\etabar_1 (D+m)^{-1}\eta_1\right\} \nonumber \\
&=&
\int \cd\ca\;e^{-S_g}\int \cd q\cd\qbar\;
\Exp\left\{-\qbar (D+m)q +\etabar_1 q +\qbar\eta_1\right\} ,
\end{eqnarray}
where $q$ is the field of the one-flavor theory.

Relaxing the restriction on the sources,
many more correlation functions can be generated by taking
derivatives of $Z(\eta,\etabar)$ with respect to all $\eta_i$ and $\etabar_i$.  The full set
thus contains many unphysical correlation functions.
Indeed, as we demonstrate in the next section,
some of these diverge in the massless limit.
This is, in fact, a generic feature of partially quenched theories,
but is fully compatible with the existence of a physical subspace,
as long as valence and sea-quark masses are equal, as they are here.

Away from the continuum limit (\ie, in actual simulations),
one has only the lattice staggered field at one's disposal,
and mixing between different tastes is unavoidable.  One method
to obtain correlation functions that become physical in the continuum limit
is to choose sources that project onto a single
taste in the continuum limit.
This is straightforward, given the known relation between
the lattice and continuum symmetry groups~\cite{gs}.

It will be useful for the next section to extend the representation~(\ref{genfun3})
to include
sources coupled to ghosts,
so as to utilize fully the symmetries of the partially quenched
theory.
We collect quarks and ghosts into a single field
with seven components:
\begin{equation}
\label{Psidef}
\Psi = (\psi_1,\psi_2,\psi_3,\psi_4,\phi_1,\phi_2,\phi_3) \,,
\end{equation}
and extend the index on the sources to run from $1$ to $7$, with
$\eta_{5-7}$ commuting. The generalized generating functional is then
\begin{equation}
\label{genfun4}
Z(\eta,\etabar)
=
\int \cd\ca\;e^{-S_g} \int \cd\Psi\cd\overline\Psi\;
\Exp\left\{-\sum_{i=1}^7\left(\overline\Psi_i (D+m)\Psi_i
+\etabar_i\Psi_i+\overline\Psi_i\eta_i\right) \right\}.
\end{equation}

If one is interested in mesonic observables only,
another projection onto the physical sector of the RCT is available,
which was discussed in detail previously \cite{cbrschpt,bgss}.
In the Appendix, we use this projection to study the effects of
't~Hooft vertices on mesonic correlation functions,
arriving at the same final conclusions, namely that the physical
sector of the RCT correctly reproduces one-flavor QCD.

\section{\label{'t Hooft} '\lowercase{t}~Hooft vertices}
In one-flavor QCD, instantons induce a bilinear 't~Hooft vertex~\cite{thooft}.
Let us first recall what this statement means.
In an instanton background,
$D$ has a left-handed zero mode $\psi_0=P_L\psi_0$.
Taking the center of the instanton to be at the origin,
the quark propagator is
\begin{equation}
\label{proptopone}
(D+m)^{-1}(y,x)=\frac{1}{m}\;\psi_0(y)\psi_0^\dagger(x)+\D(x,y)+O(m)\ ,
\end{equation}
where the $m$-independent term $\D(x,y)$ satisfies
$\{\D(x,y),\g_5\}=0$.
When we perform a Wick contraction in the fermion path integral, this
gets multiplied by $\Det(D+m)=m(\Det'(D)+O(m))$ where the prime
indicates the determinant with the contribution from the zero mode removed.
Hence, one has
(with spinor indices implicit and not contracted)
\begin{equation}
\langle \qbar(x) q(y)\rangle_F
= - \Det'(D)\, \psi_0^\dagger(x)\psi_0(y)+O(m)\ ,
\label{qqbar}
\end{equation}
where the subscript $F$ indicates
that only the fermionic functional integral has been performed.
This correlator is
non-vanishing in the limit $m\to 0$.  For distances
$|x-y|$ much larger than the instanton size, the instanton's contribution,
together with a similar contribution coming from an anti-instanton,
can be reproduced by an insertion of an effective chiral-symmetry breaking
vertex, the 't~Hooft vertex.
In the one-flavor theory it is a fermion bilinear proportional to $\qbar q$.

In the partially quenched representation of the RCT, Eq.~(\ref{genfun4}),
this result is reproduced as follows.
In view of Eq.~(\ref{determinant}), the fermion determinant in the one-instanton sector
is the same as in the one-flavor theory:
\begin{equation}
\label{detroot}
\Det^{1/4}\left[(D+m)\otimes\I\right]
=m\left(\Det'(D)+O(m)\right).
\end{equation}
The general bilinear expectation value in a fixed background gauge field
becomes
\begin{equation}
\label{PsiPsibar}
\langle\overline\Psi_i(x) \Psi_j(y)\rangle_{F}
= -\epsilon_i \d_{ij}\, \Det'(D)\, \psi^\dagger_0(x)\psi_0(y)+O(m)\ ,
\end{equation}
where the subscript $F$ now
indicates integration over quarks and ghosts.
The factor of
\begin{equation}
\label{epsilon}
\epsilon_j \equiv
\begin{cases}
  +1,& \text{if $ 1\le j\le 4$ (valence quarks);} \\
  -1,& \text{if $ 5\le j\le 7$ (ghosts).}
\end{cases}
\end{equation}
arises from the different statistics of quarks and ghosts.
Finally,
projecting onto the physical subspace by setting $i=j=1$,
one recovers the one-flavor QCD result Eq.~(\ref{qqbar}).

We next consider correlation functions containing two fermions (or ghosts
in the partially quenched theory) and
two antifermions (or antighosts).
In the one-flavor theory, no quadrilinear 't~Hooft vertices
appear.  The reason is simple. In the one-instanton sector,
any contribution in which all four fermion operators are
saturated by the (single) zero mode will vanish by Fermi statistics.
Moreover, because $\D(x,y)$ anticommutes with $\g_5$,
in the massless limit all correlation functions must violate
axial charge conservation precisely by two units, as required by the anomalous
divergence of the axial current or, equivalently, by the index theorem.

The situation in the RCT is quite different.
It is simple to see that, in a single instanton (or arbitrary topological charge $\pm 1$)
background
\begin{eqnarray}
\nonumber
\lefteqn{\langle\overline\Psi_i(x)\Psi_j(y)\overline\Psi_k(z)\Psi_l(w)\rangle_F =}
\\
&& \frac{1}{m}\, \Det'(D)\,
\psi_0^\dagger(x)\psi_0(y)\psi_0^\dagger(z)\psi_0(w)
\left( \epsilon_i\epsilon_k\d_{ij} \d_{kl} - \epsilon_i \d_{il} \d_{kj} \right)
 + O(1)\ .
\label{fourpoint}
\end{eqnarray}
Thus there is, in general, a quadrilinear effective vertex, and its
coefficient diverges like $1/m$ in the chiral limit.
If we project onto the
physical sector by setting $i=j=k=l=1$,
then the infrared singular contribution from zero-modes vanishes.
This is just the cancellation between the two fermion contractions
dictated by Fermi statistics, and reproduces the result
in one-flavor QCD.
However, in
unphysical correlation functions we can set the indices to different values.
With $i=j\ne k=l$, for example, there will be a contribution
only from one of the contractions, and the cancellation
cannot occur.
Thus there are additional vertices in the rooted theory,
with coefficients that diverge in the chiral limit.

In fact, this
is an example of a phenomenon present in any partially quenched theory.
Infrared-singular contributions, coming from the
propagators, may not be canceled by the positive powers of the quark mass,
coming from the determinants, because there is no fermion determinant
associated with any of the valence quarks.
Indeed, singularities of arbitrarily high order can occur since,
unlike for fermions, there is no limit to the number of ghost
zero-modes that can contribute.
The key point, however,
is that this pathological infrared behavior cannot take place
in the physical sector of the partially quenched theory, because this
sector does not know about the valence quarks and ghosts.
In the case of the rooted one-flavor theory,
projecting onto a single taste avoids all the singular vertices.

\section{\label{wi} Ward identities}
While the one-flavor theory and the partially quenched theory~(\ref{genfun3})
share the same physical (sub)space, they differ in their symmetries.
In the one-flavor theory, the single chiral symmetry is anomalous.
But when the one-flavor theory is embedded into the partially quenched
theory, the fermion condensate $\langle\qbar q\rangle$
transforms under some of the
non-anomalous chiral symmetries of this extended theory.
We have already demonstrated
that the fermion condensate in the partially quenched
theory,
$\svev{\psibar_1(y) \psi_1(y)}$,
 takes the same, non-zero, value as that in one-flavor QCD.
In this section we explain how this result is reconciled with
the chiral symmetries of the partially quenched theory.
Throughout this section we work in finite (though arbitrarily large)
volume so that the topological charge is well defined
(and takes integer values).

For $m>0$,
the anomalous $U(1)_A$ Ward identity of the one-flavor theory
takes the form
\begin{eqnarray}
  \svev{\d\co} &=& \svev{\d S_F\; \co} - 2\svev{Q\;  \co}
\NON
   &=& 2m \int d^4x \svev{\qbar(x) \g_5 q(x) \; \co} - 2\svev{Q\;  \co} \ ,
\label{dO}
\end{eqnarray}
where $S_F$ is the fermion action, and the topological charge
$Q = (16\p^2)^{-1} \int d^4x\, \tr (F\tilde{F})$
arises from the variation of the measure.

The Ward identity~(\ref{dO}) is in fact valid after the integration over
the fermions only.  This can be used to recover the index theorem \cite{COLEMAN}.
Let us consider a fixed gauge-field background
and choose $\co=1$, in which case the left-hand side of Eq.~(\ref{dO}) is zero.
For a fixed background field and for any $m>0$, the fermion determinant is a
non-zero common factor that may be divided out.
The index
theorem then follows by
taking the massless limit
of Eq.~(\ref{dO}):
\begin{equation}
{\rm Ind}(D) \equiv  \lim_{m\to 0} \int d^4x\;
  \Tr \left( m\g_5 (D+m)^{-1}(x,x) \right) = - Q \ .
\label{index}
\end{equation}
This expression receives contributions from zero modes only
(as follows in the case of a single instanton
from Eq.~(\ref{proptopone})).
It is an important reminder that the massless limit must be taken
carefully in order to reproduce the index theorem.

Returning to
Eq.~(\ref{dO}), we choose $\co=\qbar(y) \g_5 q(y)$ and
take the limit $m\to0$.
The first term on the right-hand side  then drops  out, since it has
two factors of $m$ in the numerator, one coming from $\d S_F$, and the other
from the fermion determinant.
(By Fermi statistics, saturating all four fermion operators with zero modes to obtain
two powers $m$ in the denominator is not possible -- see Sec.~\ref{'t Hooft}.)
We thus obtain
\begin{equation}
  \svev{\qbar(y) q(y)}
  = - \svev{Q\;  \qbar(y) \g_5 q(y)} \ .
\label{pbp}
\end{equation}
This is the expected result, with $\svev{\qbar_R q_L}$
($\svev{\qbar_L q_R}$) receiving a non-zero contribution
from the one-instanton (one anti-instanton) sector only.
We stress that the contribution is non-vanishing in the chiral limit,
even though we are in finite volume.

We next examine the expectation value of $\psibar_1(y) \psi_1(y)$
in the partially quenched theory,
and show how a non-vanishing expectation
value is consistent with the Ward identities for the
chiral symmetries of this theory.
As a concrete example we consider the symmetry
generated by $\g_5 \otimes \X_5$,
where $\X_5$ denotes a non-singlet taste generator that we may
choose as
\begin{equation}
\label{Xi5}
\X_5={\rm diag}(1,1,-1,-1)\ .
\end{equation}
Since ghost fields do not transform, we
revert to the notation in which indices run from $1$ to $4$.
This symmetry is of particular interest, since
it remains valid for valence staggered fermions even
away from the continuum limit (up to breaking by the mass term).
It is known as the $U(1)_\e$ symmetry in the staggered theory, and we keep that
nomenclature here.

Instead of Eq.~(\ref{dO}), the general $U(1)_\e$ Ward identity is
\begin{equation}
  \svev{\d\co} = 2m \sum_{i=1}^4\int d^4x
  \svev{\psibar_i(x) \, (\g_5 \otimes \X_{5,ii}) \, \psi_i(x) \; \co} \ .
\label{dOnona}
\end{equation}
Notice the absence of a ``$Q$'' term,
because the $U(1)_\e$ symmetry is not anomalous.
We now choose $\co=\psibar_1(y) \g_5 \psi_1(y)$, leading to
\begin{equation}
  \svev{\psibar_1(y) \psi_1(y)} = m \sum_{i=1}^4\int d^4x
  \svev{\psibar_i(x) \, (\g_5 \otimes \X_{5,ii}) \, \psi_i(x) \;
  \psibar_1(y) \g_5 \psi_1(y)} \ .
\label{msls0}
\end{equation}
As for Eq.~(\ref{pbp}), only $|Q|=1$ sectors can contribute to
the right-hand side in the massless limit. For there
to be a non-zero contribution in this limit
all four fields must be saturated by zero modes,
as in the leading term on the right-hand side of Eq.~(\ref{fourpoint}).
Using this result, we see that the contribution from $i=1$ vanishes,
while that from $i=2-4$ does not. Furthermore, the latter
contribution comes only from the contraction of $\psibar_i(x)$ with $\psi_i(x)$
(and $\psibar_1(y)$ with $\psi_1(y)$), so that the integration over $x$
is proportional to the index, as in~(\ref{index}).
The final result is that, in the massless limit,
\begin{equation}
  \svev{\psibar_1(y) \psi_1(y)}
  = -\sum_{i=2}^4 \X_{5,ii} \svev{{\rm Ind}(D) \; \psibar_1(y) \g_5 \psi_1(y)}
  = - \svev{Q\; \psibar_1(y) \g_5 \psi_1(y)}\ ,
\label{msls}
\end{equation}
where we have used $\sum_{i=2}^4 \Xi_{5,ii} = -1$.
This indeed takes the same form as the Ward identity
in the one flavor theory, Eq.~(\ref{pbp}).

If we integrate only over fermion and ghost fields (but not over
gauge fields) then we can extend the considerations above to the multi-local quantities
$\svev{\qbar(x) q(y)}$ and $\svev{\psibar_1(x) \psi_1(y)}$.
By an essentially identical
argument to that given above one finds that the Ward identities
from $U(1)_A$ and $U(1)_\e$ symmetries are consistent.
Thus, for example,
the presence of a bilinear 't~Hooft vertex in the partially quenched
theory is consistent with the presence of the $U(1)_\e$ symmetry, and indeed
with all of the exact (non-singlet) chiral symmetries.

We conclude from the preceding discussion that order parameters
for non-anomalous symmetries, such as
$\langle \overline\psi_i  \psi_i\rangle$,
can be non-vanishing in finite volume. This is allowed as long
as the corresponding symmetry acts in the full RCT rather than
the physical subspace. 
The essential technical point is that one cannot work directly
at $m=0$, even in finite volume, once one moves out of
the physical one-flavor subspace into the full RCT (as is required
to discuss, for example, the $U(1)_\e$ symmetry).
This is because of the severe divergences as $m\to 0$, such
as those in the quadrilinear vertex, Eq.~(\ref{fourpoint}).
Since $m\ne 0$, chiral symmetries are explicitly broken, and
there is no mathematical inconsistency in having $\svev{\psibar_i\psi_i}\ne 0$
in finite volume.

We stress
that the preceding discussion applies to the
continuum limit of the rooted staggered theory
provided that this limit is taken before
the chiral limit $m\to 0$.
This is explained in detail in Refs.~\cite{bgss,sharpelat06}.
If on the contrary, the chiral limit is taken first,
while keeping $a\ne 0$, the finite-volume condensate
will vanish \cite{dh1,sharpelat06}.
The emergence of
a non-zero chiral condensate in the Schwinger model with rooted
staggered fermions, including the non-commutativity of
the continuum and chiral limit, was carefully checked numerically
in Ref.~\cite{dh1,dh2}.

\section{\label{MC} Consequences for the argument of Ref.~[26]}
The validity of the rooted staggered theory has been called into question
by Creutz~\cite{creutz1,creutz2,creutz}, who discusses various paradoxes
that, it is claimed, provide proof of the failure of rooting.
While the discussion of Ref.~\cite{bgss}
in fact resolves all the apparent paradoxes, the most recent work
by Creutz \cite{creutz2,creutz} brings up
features of rooted theories that were not treated explicitly in Ref.~\cite{bgss}.
In particular,  the 't~Hooft vertices
of the rooted one-flavor theory are claimed in Refs.~\cite{creutz2,creutz}
to have a different structure from
the desired 't~Hooft vertices of the standard one-flavor theory.
In this section, we review each of the claims
of Refs.~\cite{creutz2,creutz} in turn, and, using the results obtained above, refute them.

\subsection{\label{racemic} A racemic mixture?}
Creutz notes \cite{creutz2,creutz} that the exact chiral symmetry of staggered
quarks (the $U(1)_\e$ symmetry) is a non-singlet symmetry in which two tastes
transform with positive chirality, and two, with negative chirality.  He calls
the resulting rooted staggered quark a ``racemic mixture''---a heterogeneous
combination of the two types of chirality. So far, this is
just a standard fact about staggered quarks, coupled with new nomenclature.
However, he then claims that the RCT of \seneqs{conlim}{genfun} cannot be the continuum
limit of a rooted staggered quark because it contains four identical tastes with
(necessarily) the same chirality, instead of the required racemic mixture.
So, for example, the RCT is required to have zero modes with the same chirality
for each taste, while the staggered quark has zero modes with opposite chiralities
for different tastes.

Creutz's argument in this case is based on a simple misunderstanding of the meaning of chirality.
The ``chirality'' of a taste (or flavor, or zero mode) is not a well-defined concept
before one specifies the particular chiral symmetry in question.  Thus, all four
tastes of the RCT have the same chirality with respect to the singlet (anomalous)
chiral symmetry, generated by $\gamma_5 \otimes I$.  But,
with respect to the non-singlet
chiral symmetry $U(1)_\e$, generated by $\gamma_5\otimes \X_5$,
there are two tastes with positive chirality ($\X_{5,11}=\X_{5,22}=+1$),
and two with negative ( $\X_{5,33}=\X_{5,44}=-1$).  So the RCT is every bit
as much a ``racemic mixture'' with respect to $U(1)_\e$ chiralities as the original
rooted staggered theory, and the same holds for the corresponding (approximate)
zero modes of the two theories.  Of course, the $U(1)_\e$ symmetry happens to be  exact
on the lattice (up to mass terms); while the singlet chiral symmetry and
all the non-singlet chiral symmetries other than $U(1)_\e$ are violated
by discretization effects.  Indeed, the singlet symmetry must be violated
on the lattice, or (even unrooted!) staggered quarks could not reproduce the correct
anomaly \cite{Sharatchandra:1981si}.  But the lesson is that, if one wants to
discuss 't~Hooft vertices, for which the anomalous symmetry is relevant, one
needs to consider the chirality of modes under that anomalous symmetry.
In both the RCT and for rooted staggered quarks on the lattice, the zero modes
for each of the tastes in the presence of an instanton have the same chirality
under the anomalous chiral symmetry.

\subsection{\label{eps} The one-flavor 't~Hooft vertex and $U(1)_\e$ symmetry}
The bilinear form of the 't~Hooft vertex of the one-flavor
theory, which has the form of a mass shift, is claimed in Ref.~\cite{creutz}
to be ``inconsistent with any exact chiral symmetry,'' and
in particular with the $U(1)_\e$ symmetry of staggered fermions.

Were this assertion correct, then it would indeed indicate a failure
of rooting. In fact, we have shown in Sec.~\ref{wi} that this claim is
not correct. The Ward identities of the $U(1)_\e$ symmetry agree
with the original, anomalous Ward identities of the $U(1)_A$ symmetry
on the physical subspace, and are consistent with a bilinear 't~Hooft vertex.
Furthermore, the direct calculation of Sec.~\ref{'t Hooft} shows that
such a vertex is present in the physical subspace, obtained by
using only a single taste.

We note again that it is essential to take the continuum limit of
rooted staggered fermions before the chiral limit.

What is missed in Ref.~\cite{creutz} is the unusual nature of symmetry
breaking in partially quenched theories---non-anomalous chiral symmetries can be
broken in finite volume. To see this one must take
the massless limit
carefully, as we have shown in Sec.~\ref{wi},
since there are more severe
infrared divergences in the enlarged set of correlation functions
of the partially quenched theory.

\subsection{\label{1/m} $1/m^n$  singularities}
Anticipating the argument that one can obtain the correct 't~Hooft vertex
by using only a single taste, Ref.~\cite{creutz} points out that the
rooted theory allows an octilinear 't~Hooft vertex,
which is not suppressed by the lattice spacing,
and that, for example, gives rise to a $1 /m^3$ divergence in the
correlation function $\svev{\prod_{i=1}^4 \psibar_i(x_i) \psi_i(y_i)}_F$.
It is then claimed that ``because of this strong coupling between the
tastes, all four must be considered in intermediate states.''

We agree with the presence of such infrared divergent 't~Hooft vertices
in the rooted theory. Indeed, the quadrilinear example, which we discussed
in Sec.~\ref{'t Hooft}, plays an essential role in Sec.~\ref{wi} in showing the
consistency between the Ward identities of one-flavor QCD and its rooted
staggered extension. Furthermore, as noted in Sec.~\ref{'t Hooft}, there are in fact
multilinear vertices involving ghosts with arbitrarily high order of infrared
divergence.

The presence of such vertices is a peculiarity of the partially quenched theory,
and the relevant question is whether these
couplings between tastes impact the physical subsector.
The answer is no, as can be seen from the formulation of the partially
quenched theory given in Eq.~(\ref{genfun3}).
As we have already seen from Eq.~(\ref{genfun-1flavor}), if the only non-vanishing sources are
$\eta_1$ and $\overline\eta_1$, which are all that is needed to generate
the correlation functions in the single taste subsector,
then the generating function is exactly that of one-flavor QCD.
There are contributions from intermediate quarks of all species,
including those that are produced by the unsuppressed coupling between tastes
due to the ``problematic'' 't~Hooft vertices.
There are also, however, contributions from intermediate ghosts,
and these precisely
cancel those from the additional tastes.
It is essential for this cancellation that one keep $m>0$ in the partially
quenched theory,  so that none of the 't~Hooft vertices actually diverge.
Only when one focuses on the physical subsector alone
(and only after taking the continuum limit of the rooted theory)
can one take the chiral limit.

The problem with the argument of Ref.~\cite{creutz} is thus seen to
be that, in the partially quenched representation, it leaves out the contribution from ghosts.
Alternatively, using the approach in the Appendix, which works directly
with the rooted theory, we can say that the problem with the argument of Ref.~\cite{creutz}
is that it does not properly take into account the cancellations that occur because
different contractions in the rooted theory are weighted with different factors of $1/4$.

We stress, as already noted in the introduction, that the additional
multilinear 't~Hooft vertices do not provide an obstacle to taking
the continuum limit of the rooted staggered lattice theory. In particular, these vertices
do not break the {\em vector} taste symmetry, and thus their
presence is consistent with the assumed partially quenched continuum limit, namely the RCT,
which has this symmetry.

\subsection{\label{IIbar} The $F\tilde{F}$ two-point function}
A specific example of the claimed additional unphysical contributions
is considered in Ref.~\cite{creutz}. This is
the two-point function of the gluonic operator
$\tr(F(x)\tilde{F}(x))$, which probes the interaction between
instantons and anti-instantons.
The specific claim is that the octilinear 't~Hooft vertices will produce
a $1/m^6$ divergent behavior in this two-point function.
Such a contribution would be clearly unphysical, and represent a failure
of rooting. As we now show, however, this contribution is absent.

The general argument based on Eq.~(\ref{genfun3}) is even more simple
than that of the previous subsection. Here we can set all fermionic
(and ghost) sources to zero, so that the partition function
(including gluonic sources as needed)
reduces to that of the one-flavor theory (with only gluonic sources).
Thus, in particular,
the correct two-point function of $F\tilde{F}$ will be reproduced.

Although this cancellation is trivial,
it is illuminating to work out the details in an instanton---anti-instanton
background.
Let us start with the physical one-flavor theory.
The instanton and anti-instanton
are located at $x$ and $y$ respectively, and both have fixed sizes.
The effective instanton---anti-instanton
interaction induced by the (nearly) massless fermions, $V(x-y)$,
is equal, by definition, to the (renormalized) fermion determinant in the
instanton---anti-instanton background:
\begin{equation}
  V_F(x-y)[\mbox{1-flavor}] = \Det(D(I,\Ibar)) \ ,
\label{int}
\end{equation}
where $D(I,\Ibar)$ denotes the single-quark Dirac operator in the background
field.
Let us \textit{assume} that $V(x-y)$
is accounted for by the 't~Hooft vertices.
Since these vertices are
$H_I \propto \psibar_R \psi_L$ and $H_{\Ibar} \propto \psibar_L \psi_R$,
we have
\begin{equation}
  V_F(x-y)[\mbox{1-flavor}]
  = \svev{H_I(x) H_\Ibar(y)}\;
  \propto\; 1/(x-y)^6\ ,
\label{int1}
\end{equation}
where
the fermion contractions are performed in the free theory
(as is always the case---by definition---with 't~Hooft vertices.)
Extending this result to the \textit{unrooted} continuum theory
we evidently have
\begin{equation}
  V_F(x-y)[\mbox{unrooted}]
  = \Det^4(D(I,\Ibar))
  \propto\; 1/(x-y)^{6\times 4}\ ,
\label{int4}
\end{equation}
whereas for the RCT, which is equivalent to the partially quenched
$(4|3)$ theory,
\begin{equation}
  V_F(x-y)[\mbox{rooted}]
  = \Det^{4-3}(D(I,\Ibar))
  \propto\; 1/(x-y)^{6\times (4-3)}\ .
\label{introot}
\end{equation}
As expected, this agrees with the one-flavor result.
This conclusion is, clearly, completely independent of the detailed
form of the instanton---anti-instanton effective interaction \cite{IIb}.

It is now worth noting
that the effective interaction induced by a single \textit{ghost}-quark is
\begin{equation}
  V_F(x-y)[\mbox{one ghost}]
  = \Det^{-1}(D(I,\Ibar))
  \propto\; (x-y)^6\ .
\label{ghost}
\end{equation}
The ghost-quark induced interaction is growing
with the separation!  Once again Eq.~(\ref{ghost}) is nothing more than
a trivial consequence of the fact that the ghost-quark determinant
is by construction the inverse of the quark determinant.
Nevertheless, this implies that the ghost-quark contribution
is not amenable to the language of 't~Hooft vertices; the interaction
between local operators never grows with distance.
In this language, the erroneous conclusion of Ref.~\cite{creutz} (see in particular Fig.~5 therein)
is a result of failing to take into account the contribution of the ghost
quarks.

\subsection{\label{motion} Motion of eigenvalues between topological sectors}
For most gauge fields near the continuum limit, the eigenvalues
$\l_i$ of $D_{\rm stag}$ with $a|\l_i|\ll 1$
are observed to lie in approximate quartets,
consistent with the approximate taste symmetry~\cite{Durrev,evs}.
The  softly broken $U(1)_\e$ symmetry plays no role in determining
the approximate quartet structure, but does imply
that all eigenvalues appear in pairs with opposite imaginary parts,
$\pm i \lambda + m$.
The number of quartets with $\l$ near zero matches,
for most configurations, the number expected from the index theorem.
Thus, for example, a configuration with unit topological charge has
a ``zero-mode quartet'' composed of two $U(1)_\e$ pairs: there
are two eigenvalues with (small) positive imaginary parts and two with
the corresponding negative imaginary parts.

It is observed in Ref.~\cite{creutz} that the quartet structure cannot
be maintained as one traverses from one topological charge sector to another
(as one can do on the lattice by varying the gauge field continuously).
For example, moving from $Q=0$ to $Q=1$, a quartet of approximate zero modes
must appear, and this can only happen by having two eigenvalues with
positive imaginary part come down to the real axis, with their $U(1)_\e$
partners coming up symmetrically from below. Thus at least two approximate
quartets must be broken up during the transit (one each for positive and
negative imaginary part). This is indicated pictorially in Figure 3 of
Ref.~\cite{creutz}.

In this case we agree with the description of Ref.~\cite{creutz},
but argue that it does not present a problem for rooting.
It is true that the transit between sectors involves rough gauge fields
with significant components with momenta $p\sim 1/a$,
and this does lead to significant taste breaking.
The issue, however, is whether such gauge fields are important in the
functional integral.  In the continuum theory, different topological
sectors form disconnected spaces.
We thus expect that, regardless of the type of fermions used,
lattice configurations lying ``on the boundary'' between different
topological sectors must have a vanishing
weight in the continuum limit.

\section{\label{conclusion} Conclusion}
In this paper, we have
shown how the appropriate 't~Hooft vertices
appear in a rooted version of QCD with exact taste symmetry,
the so-called rooted continuum theory (the ``RCT''),
and how they are consistent with the Ward identities of the rooted theory.
A key to doing so is realizing that the RCT
is a partially quenched theory.
A crucial property of such theories is that they contain a physical sector
(here the desired target theory, QCD)
that is protected from the unphysical effects that are present in the
full partially quenched theory.
It is straightforward to construct correlation functions that are contained
in the physical sector of the RCT, and thus are free from unphysical effects.
In the main text we used a projection onto
a single taste, while in the Appendix we show how a different
projection, used in our earlier paper~\cite{bgss}, also works.
For each of these approaches, there exist lattice versions that go over
to the desired projections in the continuum limit and that
can easily be applied to the rooted staggered theory.

A subsidiary result of this paper is to expose further unphysical features
of partially quenched theories.
That unphysical effects are present is well known.
What we find, elaborating on an observation of Ref.~\cite{creutz},
is that (anti-)instantons
give rise to fermionic correlation functions that diverge
with a power of $m$ when $m\to 0$. We observe that such
infrared divergent 't~Hooft vertices exist also with external ghost quarks,
and that for these there is no limit to the power of the divergence.
Nevertheless, these quenched sicknesses do not affect the physical sector,
because of a cancellation between valence quark and ghost sectors.

Another unphysical feature is that 
an order parameter for a non-anomalous symmetry
of a partially quenched theory can be non-vanishing
in finite volume, if the action of the symmetry is not restricted
to the physical subspace.
This can happen because
the divergences in correlation functions force one to approach the
massless limit carefully from $m\not=0$, where chiral symmetries are explicitly broken.
This phenomenon occurs in rooted one-flavor QCD, and plays a
central role in understanding how the condensate of that theory
is consistent with the extended symmetries of the RCT, and
hence also of rooted staggered QCD.

Our discussion was mostly restricted to the continuum limit, and our
observations do not imply that rooted staggered fermions have been
proven to be correct.  What they do imply is that it is not possible to
invalidate rooting using arguments based on the symmetries of the
partially quenched theory that is the (assumed)
continuum limit of rooted staggered QCD. Indeed, because of
discretization effects, the symmetries of rooted staggered QCD are a subset
of those of the RCT. So if the RCT does not have what
Creutz calls ``too much symmetry'' \cite{creutz2} to
preclude it having a physical subspace equivalent to QCD (as we have shown), then
the rooted staggered theory itself cannot have ``too much symmetry.''
We have made this point previously in Ref.~\cite{bgss} in the context
of mesonic correlation functions. What we have done here extends the
discussion to all correlators.
In particular, our
arguments invalidate the claims of Refs.~\cite{creutz1,creutz2,creutz} that rooting fails.
The core claim of Ref.~\cite{creutz} is that the appropriate 't~Hooft vertices
cannot occur (\eg, a bilinear vertex in one-flavor QCD) because
they are inconsistent with the symmetries of staggered fermions.
On the contrary, we show that
the proper 't~Hooft vertices exist provided that the continuum limit
is taken before the chiral limit, and that they are consistent with the
staggered symmetries.

\vspace{3ex}
\noindent {\bf Acknowledgments}
\vspace{3ex}

We thank Andreas Kronfeld for discussions.
CB, MG and SS were supported in part by the US Department of Energy.  YS was
supported by the Israel Science Foundation under grant no.~173/05.

\appendix
\section{\label{appendix} An alternative projection}
%
In this Appendix, we discuss an alternative projection onto the physical
subspace of the RCT. This projection was introduced in Ref.~\cite{bgss}, to which
we refer the reader for more details, and
works only for mesonic correlation functions.
It is a useful  projection to consider, because it can be easily applied
to staggered fields on the lattice.  In addition, it does not require
the introduction of ghosts for the treatment of the 't~Hooft vertices.

For mesonic operators, one may introduce a source $J$ in the rooted determinant:
\begin{equation}
\label{detsource}
\Det^{1/4}\left(D_{RCT}+M +J\right)= \Det^{1/4}\left((D+m)\otimes\I+J\right)\ .
\end{equation}
Restricting to a taste-singlet source, $J=\tJ\otimes\I$, we have
\begin{equation}
\label{root}
\Det^{1/4}\left(D_{RCT}+M +\tJ\otimes\I\right)= \Det^{1/4}\left((D+m+\tJ)\otimes\I\right)
=\Det\left(D+m+\tJ\right)\ .
\end{equation}
The rightmost expression can be used to generate the set of
mesonic correlation functions of the physical one-flavor theory; therefore
the same is true for the source $J=\tJ\otimes\I$ in the rooted theory.
However, if we generate correlation functions using the
leftmost or middle expression, both of which include explicitly the taste
degrees of freedom,
we will get intermediate states including
all combinations of tastes,
as well as factors of $1/4$ coming from
the fourth root of the determinant. The factors  will depend on the particular
contraction in question; it is in fact easy to see that we get one power of $1/4$
for each quark loop.  Thus Eq.~(\ref{root}) implies that the factors of $1/4$ must
compensate for the presence of intermediate states made out of all tastes,
giving precisely the physical one-flavor mesonic correlations.

Let us work out two examples relevant to the discussion of 't~Hooft
vertices in Sec.~\ref{'t Hooft}.  Comparing $\tJ$ derivatives of the
rightmost and leftmost expressions in Eq.~(\ref{root}), we find, in
the one-flavor theory:
\begin{subequations}
\label{example}
\begin{eqnarray}
\langle\qbar(x)q(x)\rangle_F&=&
-\frac{1}{4}\, \Det^{1/4}(D_{RCT}+M)\, \Tr\left((D_{RCT}+M)^{-1}(x,x)\right) , \hspace{5ex} \label{examplea}\\
\hspace{-0.6cm}\langle\qbar(x)q(x)\;\qbar(y)q(y)\rangle_F
&=& \Det^{1/4}(D_{RCT}+M)
\label{exampleb}
\\
&&\hspace{-1.7cm}
\times\bigg[ -\frac{1}{4}\,\Tr\left((D_{RCT}+M)^{-1}(x,y)\;(D_{RCT}+M)^{-1}(y,x)\right)\nonumber\\
&&\hspace{-1.7cm}
+\,\frac{1}{16}\,\Tr\left((D_{RCT}+M)^{-1}(x,x)\right)\Tr\left((D_{RCT}+M)^{-1}(y,y)\right) \bigg]\;,
\nonumber
\end{eqnarray}
\end{subequations}
where again $q,\qbar$ are the one-flavor fields, and $\langle\dots\rangle_F$ denotes averaging over fermion fields only.
As expected, different contractions are weighted by different powers of $1/4$ on the
right-hand side.
The corresponding lattice expressions can be obtained simply by replacing
$D_{RCT}+M$ by $D_{stag} + m_{stag}$.

With exact taste symmetry, each trace in Eq.~(\ref{example}) produces a factor of $4$,
canceling the factors of $1/4$.  Thus, for example, the two terms in Eq.~(\ref{exampleb})
end up with equal and opposite weights.
When saturated by zero modes
in the $Q=1$ sector, each of these diverges as $1/m$, but
this divergence cancels in the sum.
This cancellation corresponds exactly to the cancellation
among the two terms in Eq.~(\ref{fourpoint}) when that equation
is projected on the physical subspace
by choosing $i=j=k=l=1$.
As explained above, this is a manifestation of Fermi statistics.
It explicitly shows that the unphysical four-point
't~Hooft vertex does not appear in the physical subspace of the rooted theory.

On the other hand, if we take derivatives with respect to the more general sources in
Eq.~(\ref{detsource}), we can generate unphysical correlations.  For example,
taking derivatives with respect to $J_{11}$ and $J_{22}$,
and using the exact taste symmetry, gives
\begin{eqnarray}
\hspace{-0.5cm}
\frac{\delta^2\;\Det^{1/4}\left(D_{RCT}+M +J\right)}{\delta J_{11}(x)\; \delta J_{22}(y)}
\Bigg\vert_{J=0} &=&  \Det^{1/4}(D_{RCT}+M)
\label{unphysical}\\
&&\hspace{-3.5cm} \times
\frac{1}{16}\,\Tr\left([(D_{RCT}+M)^{-1}]_{11}(x,x)\right)\Tr\left([(D_{RCT}+M)^{-1}]_{22}(y,y)\right).
\nonumber
\end{eqnarray}
There is no cancellation here, and there will be a $1/m$ divergence in this
correlator in the $Q=1$ sector coming from zero modes.  This is as expected,
since this correlation function is outside the physical subspace.
We see again that in the rooted theory one has to take
$m\ne 0$, and that the limit $m\to 0$ can only be taken in the
physical subspace.

\end{document}